\documentclass
{mn2e}
\usepackage{epsfig}
\usepackage{amsmath, amssymb,bm}

\def \be{\begin{equation}}
\def \ee{\end{equation}}

\def \msun{\rm M_{\odot}}

\def  \le{{L_{\rm Edd}}}

\begin{document}
%%%%%%%%%%%%%%%%%%%%%%%%%%%%%%%%%%%%%%%%%%%%%%%%%%%%%%%%%%%%%%%%%%%%%%%%%%%%%%
%% Title Details and Page Header                                            %%
%%%%%%%%%%%%%%%%%%%%%%%%%%%%%%%%%%%%%%%%%%%%%%%%%%%%%%%%%%%%%%%%%%%%%%%%%%%%%%
\title[] {Ultramassive Black Holes and the Three M -- Sigma Relations}

\author[Andrew King] 
{\parbox{5in}{Andrew King$^{1, 2, 3}$ 
}
\vspace{0.1in} \\ $^1$ School of Physics \& Astronomy, University
of Leicester, Leicester LE1 7RH UK\\ 
$^2$ Astronomical Institute Anton Pannekoek, University of Amsterdam, Science Park 904, NL-1098 XH Amsterdam, The Netherlands \\
$^{3}$ Leiden Observatory, Leiden University, Niels Bohrweg 2, NL-2333 CA Leiden, The Netherlands}

\maketitle

\begin{abstract}
I consider recent observations of ultramassive black holes. These appear to confirm theoretical predictions that the relation between central black hole mass $M$ and spheroid velocity dispersion $\sigma$ has the same form $M \propto \sigma^4$ in spiral galaxies, elliptical galaxies, and cluster ellipticals, but has differing normalizations. These arise from the need for longer black hole accretion episodes to expel the gas otherwise potentially able to feed the holes in the latter two types of host. In a sample drawn from a mixture of galaxy host types the fitted power of $\sigma$ will slightly exceed the theoretically--derived value of 4 because of the differing normalizations. 
The observed hole masses do not currently reach the theoretical maximum values possible for disc accretion, set by the equality of the ISCO and self--gravity radii, probably because the host galaxies have insufficient gas.  
\end{abstract}

Comments: Accepted for MNRAS


\footnotetext[1]{E-mail: ark@astro.le.ac.uk}
\section{Introduction}
\label{intro}
In a recent paper, De Nicola et al (2025; hereafter DN25) consider ultramassive black holes (UMBHs), defined as having masses $M > 10^{10}\msun$. In a sample of 16 Brightest Cluster Galaxies (BCGs) with no previous black--hole mass measurements, they find 8 UMBHs from direct dynamical detections with triaxial Schwarzschild models. When plotted on the standard $M - \sigma$ plane relating black hole mass to galaxy bulge velocity dispersion $\sigma$ these UMBHs form a sequence parallel to, but above, the standard $M - \sigma$ relation (see Fig 1 of DN25).

I argue here that these observations accord with theoretical predictions for the central black hole masses in BCGs, which are all ellipticals.  The normalization of the $M - \sigma$ relation here is significantly larger than for spiral galaxies, because black--hole feedback has to expel gas from a bigger spheroid before SMBH growth is finally halted.

%The observed UMBH masses all have the property that the radius where self--gravity fixes the outer edge of any accretion disc 
%is only slightly larger than the ISCO radius specifying the disc's inner edge, implying that 
%
%these are close to the largest black hole masses that can have grown through normal disc accretion in these galaxies. 
%
\section{Scaling Relations for Black Holes and Host Galaxies}

A wealth of observations show that the masses $M$ of supermassive black holes (SMBH) are closely related to properties of their host galaxies in two ways, i.e.
%\be
%M   = \frac{f_g\kappa}{\pi G^2}\sigma^4 \simeq 3\times 10^8\msun\sigma_{200}^4
%\label{Msig}
%\ee
%(where $f_g is the gas mass fraction relative to stars near the hole, and $\kappa$ is the electron--scattering opacity), which implies
\be
M \simeq M_{\sigma} \simeq 3\times 10^8\sigma_{200}^4\msun
\label{msigobs}
\ee
 (Ferrarese \& Merritt, 2000; Gebhardt et al. 2000), and 
\be 
M \simeq 10^{-3}M_b.
\label{Mm}
\ee
Here $\sigma$ is the central velocity dispersion of the host spheroid, with $\sigma_{200} = \sigma/(200\, {\rm km\, s^{-1})}$, and $M_b$ is the
total spheroid stellar mass (see Kormendy \& Ho, 2013 for references and a review).

A widely--accepted interpretation (see King \& Pounds, 2015 and King, 2023 for reviews) is that the first relation (\ref{msigobs}) results from the direct Eddington--limited momentum--driven feedback. Momentum--driving is established because the shocks of the 
radiation--pressure--driven black--hole outflows against the host interstellar gas are cooled by the inverse Compton effect of the active galaxy nucleus's radiation field. The $M - \sigma$ mass is fixed at the point where the shocks expand to radii $R_C \sim 1$~kpc so large that the radiation field is unable to maintain sufficient inverse Compton cooling, and the flow changes from momentum--driven to energy--driven, sweeping the surrounding gas away at high speed\footnote{Note that the assumption of energy--driving from the outset (e.g. Silk \& Rees, 1998) predicts black hole masses far smaller than the observed values (see King \& Pounds, Section 7.2, or King, 2023 Section 6.11}. 
Once the hole mass $M$ reaches the $M - \sigma$ value
\be
M_{\sigma} =\frac{f_g\kappa}{\pi G^2}\sigma^4 \simeq 3\times 10^8\sigma_{200}^4\msun
\label{msig}
\ee
(where $f_g$ is the gas fraction by mass of the spheroid and $\kappa\simeq 0.34$ is the electron scattering opacity),
 its feedback is able to expel the spheroid gas, 
 %%%
 %% 
 % because Compton cooling by the radiation field from the SMBH vicinity is sharply weakened,
 %  causing the outflow to become energy--driven. This inhibits and ultimately prevents
 %%%
                  inhibiting and perhaps ultimately preventing 
 %%%
any further black--hole mass growth (King, 2003; 2005).  
%%%
This picture explains observations of AGN with momentum--conserving flows close to the black  hole driving high--speed molecular outflows on large scales (Tombesi et al., 2015). Recent 
spatially--resolved observations (Marconcini et al., 2025) explicitly show that the outflows do indeed accelerate sharply at the radius $\sim 0.5 - 1\,{\rm kpc}$ where the Compton cooling time becomes longer than the flow time.
 %%%%%%%%%%%%%%%%%%%%%%%%%%%%%
 
The second relation (i.e. $M \propto M_b$) arises because 
momentum feedback from the bulge stellar population via winds and supernovae limits the total stellar spheroid mass $M_b$ to a value which is also $\propto \sigma^4$, and so proportional to $M_{\sigma}$ (Power et al, 2011). 

Importantly, the precise normalization of the $M-\sigma$ mass ($\propto \sigma^4$) in (\ref{msig}) depends on how long it takes the black hole feedback to expel all the spheroid gas from its vicinity, finally switching off accretion. If this lasts a time $t$  the SMBH mass must grow beyond the value (\ref{msig}) as 
\be
M \simeq M_{\sigma} e^{t/t_{\rm S}}, 
\ee
where
\be
t_{\rm S} = \frac{\eta\kappa c}{4\pi G}\simeq 4\times 10^7\,{\rm yr} \simeq 4\times 10^7\, {\rm yr}
\ee
is the Salpeter time.

Zubovas \& King (2012) noted that in a spiral galaxy with bulge radius  
$\sim 1 - 2$~kpc, all of the bulge gas is removed by energy--driving
on a timescale significantly smaller than 
$t_{\rm S}$, so the SMBH mass remains at the value (\ref{msig}), in agreement with observation.
But if the spheroid is more extended, as in elliptical galaxies, the SMBH must inject more energy into the surroundings,  requiring further SMBH mass growth. 

This effect is significant in elliptical galaxies because the spheroid has a size of order the virial radius, and is even larger if the elliptical host is part of a cluster of galaxies. In particular,
the results found by DN25 show that UMBHs closely follow the observed positive correlation between black hole mass $M$ and core radius $R_{\rm c}$ (I particularly thank the referee for this remark). This fits with the idea that the heaviest black holes result from mergers of massive galaxies. Their central SMBHs form a binary, which then ejects nearby stars via gravitational slingshots. This sets the $M - R_{\rm c}$ relation, and so determines the remaining bulge gas mass which black--hole feedback removes, so setting the normalization of the $M - \sigma$ relation.

%It would be valuable if the author could discuss whether this trend is consistent with the proposed framework of an M-sigma relation with host-dependent normalizations.

Accordingly, Zubovas \& King (2012) derived three parallel $M - \sigma$ relations, for field spirals, field ellipticals and cluster ellipticals (their Fig. 10).
All of these relations have $M \propto \sigma^4$, but with increasing normalizations: in principle there is a fourth relation for cluster spirals, but these are rare. For brightest cluster ellipticals, as studied by DN25, the time to expel all the central gas is of order two Salpeter times, and the resulting parallel $M - \sigma$ relation for bright cluster ellipticals is 
\be
M_{\sigma}({\rm BCE})\sim e^2 M_{\sigma} \simeq 7.5M_{\sigma} = 2.3\times 10^9\sigma_{200}^4\msun.
\label{BCE}
\ee

The relation (\ref{BCE}) is an extremely close fit to the BCE $M - \sigma$ values found by DN25 (the red sequence plotted in their Fig. 1),
giving for example an SMBH mass $M = 1.4\times 10^{10}\msun\sigma_{200}^4\msun$ for a velocity dispersion 
$\sigma = 300~{\rm km}\,{\rm s}^{-1}$.

\section{Relation to the Theoretical Maximum Black Hole Mass}

The UMBH masses $\lesssim 5\times 10^{10}\msun$ plotted in Fig. 1 of DN25 represent the highest SMBH masses currently known or deduced (compare with the list in King, 2016). They are found by Schwarzschild modelling, so there is no direct evidence that these masses grew by luminous gas accretion through a disc, as opposed to merging with other SMBH as a result of galaxy mergers, or swallowing stars without disrupting them. 

None of these UMBH are above the theoretical maximum black--hole mass 
\be
M_{\rm max} \simeq 
5\times 10^{10}\msun\alpha_{0.1}^{7/13}(L/L_{\rm Edd})^{-4/13}f_5^{-27/26}
\label{mmax}
\ee
permitting luminous accretion via a disc (King, 2016), where $\eta, f_5(a)$ are specified 
parametrically as functions of the spin parameter $a$ (see the relations 9 and 11 in King \& 
Pringle 2006). The current record--holder is 0014 + 813, with 
$M = 4\times10^{10}\msun$ (Ghisellini et al. 2010)

The limit (\ref{mmax}) is derived by requiring that the minimum disc radius (equal to that of the innermost stable circular orbit, or ISCO) should be smaller than the maximum radius $R_{\rm sg} \simeq 3\times 10^{15}~{\rm cm}$ (Collin--Souffrin \& Dumont, 1990; see also King et al. 2008 and King, 2023) beyond which the disc becomes 
self--gravitating and viscous accretion is not possible. (Note that $R_{\rm sg}$ is remarkably independent of all disc parameters: see King, 2023, Section 4.9 for a discussion.)

This suggests that the observed UMBH masses are simply fixed by the $M - \sigma$ relation for 
cluster ellipticals. They do not reach $M_{\rm max}$ because there is too little gas available for accretion. In principle they could exceed this value if they underwent a merger with a similarly very massive black hole, but this is evidently a very rare event.

\section{Conclusion}

I conclude that observations currently confirm theoretical estimates of the differing amounts of mass expulsion required for supermassive black holes to arrive at their final $M - \sigma$ masses  in spiral, elliptical and cluster elliptical hosts. These masses are all $\propto \sigma^4$, but have differing normalizations. 

I note that in a sample of UMBH drawn from all types of galaxy hosts, the largest hole masses correspond to ellipticals, where the normalizations vs $\sigma^4$ are largest. This may account for a tendency for the observationally--estimated exponent of $\sigma$ to exceed the theoretical value of 4.

\section*{DATA AVAILABILITY}
No new data were generated or analysed in support of this research.

\section*{ACKNOWLEDGMENTS}
I am very grateful to the referee for drawing my attention to the positive correlation of SMBH mass and galaxy core radius, as discussed above.

%Jean--Pierre Lasota for many discussions of ULX behaviour, and the referee for thoughtful comments which improved the paper.

{}
\end{document}